\documentclass[letterpaper, 10 pt, conference]{ieeeconf}  

\IEEEoverridecommandlockouts                              

\usepackage{cite}
\usepackage{amsmath,amssymb,amsfonts}
\usepackage{graphicx} 
\usepackage{amsmath}
\usepackage{amssymb}  
\usepackage{mathrsfs}
\usepackage{multicol}
\usepackage{xcolor}
\usepackage{float}
\usepackage{subcaption}
\usepackage{mathtools}
\usepackage{soul}
\usepackage{bm}
\usepackage[T1]{fontenc}
\newtheorem{remark}{Remark}
\newtheorem{assumption}{Assumption}
\newtheorem{definition}{Definition}

\newtheorem{problem}{Problem}

\newtheorem{theorem}{Theorem}

\newtheorem{lemma}{Lemma}
\usepackage{tikz}
\usepackage{xcolor}
\usepackage{dsfont}
\usepackage{algorithm}
\usepackage{algpseudocode}
\usepackage{comment}
\usepackage{hyperref} 

\let\labelindent\relax
\usepackage[shortlabels]{enumitem}

\DeclareMathAlphabet\mathbfcal{OMS}{cmsy}{b}{n}

\definecolor{darkspringgreen}{rgb}{0.09, 0.45, 0.27}

\def\BibTeX{{\rm B\kern-.05em{\sc i\kern-.025em b}\kern-.08em
    T\kern-.1667em\lower.7ex\hbox{E}\kern-.125emX}}

\def\1{\mathds{1}}

\title{\LARGE \bf
Safety-Aware Performance Boosting for Constrained Nonlinear Systems
}

\author{Danilo Saccani, Haoming Shen, Luca Furieri, Giancarlo Ferrari-Trecate
\thanks{This work was supported by the Swiss National Science Foundation (SNSF) through the NCCR Automation, a National Centre of Competence in Research (grant number 51NF40\_225155). Luca Furieri is grateful to the SNSF for the Ambizione grant (grant  number PZ00P2\_208951).}
\thanks{D. Saccani and G. Ferrari-Trecate are with the Institute of Mechanical Engineering, Ecole Polytechnique Fédérale de Lausanne (EPFL), CH-1015 Lausanne, Switzerland. (\texttt{\{danilo.saccani, giancarlo.ferraritrecate\}@epfl.ch)}. H. Shen is with the IMT School for Advanced
Studies, Lucca, Italy (\texttt{haoming.shen@imtlucca.it)}. L. Furieri is with the Department of Engineering Science, University of Oxford, United Kingdom (\texttt{luca.furieri@eng.ox.ac.uk}).%
}
}
\begin{document}

\maketitle
\thispagestyle{empty}
\pagestyle{empty}

\begin{abstract}
We study a control architecture for nonlinear constrained systems that integrates a performance-boosting (PB) controller with a scheduled Predictive Safety Filter (PSF). The PSF acts as a pre-stabilizing base controller that enforces state and input constraints. The PB controller, parameterized as a causal operator, influences the PSF in two ways: it proposes a performance input to be filtered, and it provides a scheduling signal to adjust the filter's Lyapunov-decrease rate. We prove two main results: (i) Stability by design: any controller adhering to this parametrization maintains closed-loop stability of the pre-stabilized system and inherits PSF safety. (ii) Trajectory-set expansion: the architecture strictly expands the set of safe, stable trajectories achievable by controllers combined with conventional PSFs, which rely on a pre-defined Lyapunov decrease rate to ensure stability. This scheduling allows the PB controller to safely execute complex behaviors, such as transient detours, that are provably unattainable by standard PSF formulations. 
We demonstrate this expanded capability on a constrained inverted pendulum task with a moving obstacle.
\end{abstract}


\section{Introduction}
To operate effectively in complex environments, autonomous systems must go beyond simple stabilization. The main goal is to execute advanced tasks to improve \emph{performance}, while guaranteeing \emph{safety} at all times. Reconciling these performance goals with formal proofs of closed-loop \emph{stability} and constraint satisfaction remains an open problem.
Model Predictive Control (MPC) enforces safety through hard constraints while optimizing a finite-horizon cost~\cite{rawlings2020model}. However, coupling safety, stability, and performance in a single online optimization can be conservative~\cite{saccani2022multitrajectory}. Although multi-trajectory MPC schemes partially mitigate this issue by separating safe'' and exploitation'' trajectories~\cite{saccani2023model,saccani2022multitrajectory}, they still induce an implicit policy based on a finite-horizon approximation of the underlying nonlinear optimal control problem, which may remain conservative and computationally demanding. This motivates the search for richer, explicit policy classes.
Learning-based approaches, such as Reinforcement Learning (RL), offer a promising alternative by optimizing expressive, explicit policies from data~\cite{tang2025deep}. However, generic RL pipelines often lack formal guarantees for safety and stability~\cite{gu2024review}. 
Recent works exploit Internal Model Control (IMC) structure and Youla-style parameterizations for nonlinear systems to design neural network controllers with closed-loop stability guarantees~\cite{furieriLearningBoostPerformance2024,saccani2024optimal, furieri_neural_2022,wang2022youla, galimberti2024parametrizations}. While these methods offer a clean parametrization of all and only the controllers preserving the $\ell_p$-stability properties of a given, possibly prestabilized, nonlinear system, they typically do not natively handle hard state and input constraints required for safety.
A model-based strategy to provide these guarantees is to filter unsafe inputs before application~\cite{wabersichDataDrivenSafetyFilters2023}. 
While methods like Control Barrier Functions (CBFs) can enforce safety via online Quadratic Programming (QPs), they often require handcrafted certificates and can face infeasibility issues~\cite{ames2019control}.
Predictive Safety Filters (PSFs)~\cite{WABERSICH2021109597} are MPC-based filters 
which penalizes deviations from a desired input. Recent variants add an explicit Lyapunov-decrease constraint to also enforce stability~\cite{miliosStabilityMechanismsPredictive2024,didier2024predictive}. While effective, these PSFs have two major limitations. First, the fixed, monotonic decrease constraint permanently confines the system within shrinking Lyapunov level sets. This can be overly conservative and fundamentally limits the set of achievable closed-loop trajectories, rendering complex behaviors, such as safe detours to navigate around an obstacle, impossible. Second, training a parametrized controller wrapped around a PSF with a gradient-descent approach is practically difficult, as it requires differentiating through a non-smooth optimization problem, whose solution map can be non-differentiable when constraints become active~\cite{andersson_sensitivity_2018}.

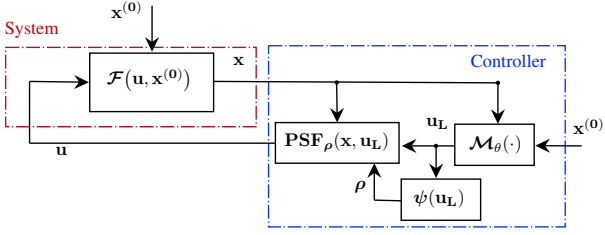
\begin{figure}[t]
    \centering
    \tikzset{every picture/.style={line width=0.75pt}} 
\resizebox{0.95\columnwidth}{!}{%

\tikzset{every picture/.style={line width=0.75pt}} 

\begin{tikzpicture}[x=0.75pt,y=0.75pt,yscale=-1,xscale=1]

\draw  [color={rgb, 255:red, 0; green, 59; blue, 255 }  ,draw opacity=1 ][dash pattern={on 7.5pt off 1.5pt on 0.75pt off 1.5pt}] (254,57) -- (470,57) -- (470,189) -- (254,189) -- cycle ;
\draw  [color={rgb, 255:red, 208; green, 2; blue, 27 }  ,draw opacity=1 ][dash pattern={on 7.5pt off 1.5pt on 0.75pt off 1.5pt}] (62.6,58.2) -- (245,58.2) -- (245,116) -- (62.6,116) -- cycle ;
\draw   (123.8,62.4) -- (214,62.4) -- (214,105) -- (123.8,105) -- cycle ;
\draw   (389.6,114.17) -- (447.4,114.17) -- (447.4,144.17) -- (389.6,144.17) -- cycle ;
\draw    (169.2,27.8) -- (169.2,58.4) ;
\draw [shift={(169.2,61.4)}, rotate = 270] [fill={rgb, 255:red, 0; green, 0; blue, 0 }  ][line width=0.08]  [draw opacity=0] (10.72,-5.15) -- (0,0) -- (10.72,5.15) -- (7.12,0) -- cycle    ;
\draw    (214,83) -- (419.9,84.04) ;
\draw    (482,130) -- (451.2,129.82) ;
\draw [shift={(448.2,129.8)}, rotate = 0.34] [fill={rgb, 255:red, 0; green, 0; blue, 0 }  ][line width=0.08]  [draw opacity=0] (10.72,-5.15) -- (0,0) -- (10.72,5.15) -- (7.12,0) -- cycle    ;
\draw    (80,128) -- (259,128) ;
\draw    (80,83) -- (80,128) ;
\draw    (80,83) -- (120.2,83.56) ;
\draw [shift={(123.2,83.6)}, rotate = 180.8] [fill={rgb, 255:red, 0; green, 0; blue, 0 }  ][line width=0.08]  [draw opacity=0] (10.72,-5.15) -- (0,0) -- (10.72,5.15) -- (7.12,0) -- cycle    ;
\draw    (389,130) -- (354,130) ;
\draw [shift={(351,130)}, rotate = 360] [fill={rgb, 255:red, 0; green, 0; blue, 0 }  ][line width=0.08]  [draw opacity=0] (10.72,-5.15) -- (0,0) -- (10.72,5.15) -- (7.12,0) -- cycle    ;
\draw    (303.6,84.77) -- (303.96,110) ;
\draw [shift={(304,113)}, rotate = 269.19] [fill={rgb, 255:red, 0; green, 0; blue, 0 }  ][line width=0.08]  [draw opacity=0] (10.72,-5.15) -- (0,0) -- (10.72,5.15) -- (7.12,0) -- cycle    ;
\draw   (259,113) -- (349,113) -- (349,142.6) -- (259,142.6) -- cycle ;
\draw  [fill={rgb, 255:red, 0; green, 0; blue, 0 }  ,fill opacity=1 ] (302.5,83.74) .. controls (302.5,83.16) and (302.99,82.7) .. (303.6,82.7) .. controls (304.21,82.7) and (304.7,83.16) .. (304.7,83.74) .. controls (304.7,84.31) and (304.21,84.77) .. (303.6,84.77) .. controls (302.99,84.77) and (302.5,84.31) .. (302.5,83.74) -- cycle ;
\draw    (421,85.07) -- (421,111) ;
\draw [shift={(421,114)}, rotate = 270] [fill={rgb, 255:red, 0; green, 0; blue, 0 }  ][line width=0.08]  [draw opacity=0] (10.72,-5.15) -- (0,0) -- (10.72,5.15) -- (7.12,0) -- cycle    ;
\draw  [fill={rgb, 255:red, 0; green, 0; blue, 0 }  ,fill opacity=1 ] (419.9,84.04) .. controls (419.9,83.46) and (420.39,83) .. (421,83) .. controls (421.61,83) and (422.1,83.46) .. (422.1,84.04) .. controls (422.1,84.61) and (421.61,85.07) .. (421,85.07) .. controls (420.39,85.07) and (419.9,84.61) .. (419.9,84.04) -- cycle ;
\draw    (331,169.5) -- (331,145.6) ;
\draw [shift={(331,142.6)}, rotate = 90] [fill={rgb, 255:red, 0; green, 0; blue, 0 }  ][line width=0.08]  [draw opacity=0] (10.72,-5.15) -- (0,0) -- (10.72,5.15) -- (7.12,0) -- cycle    ;
\draw   (350.6,154.17) -- (408.4,154.17) -- (408.4,184.17) -- (350.6,184.17) -- cycle ;
\draw    (376,131) -- (376,150.5) ;
\draw [shift={(376,153.5)}, rotate = 270] [fill={rgb, 255:red, 0; green, 0; blue, 0 }  ][line width=0.08]  [draw opacity=0] (10.72,-5.15) -- (0,0) -- (10.72,5.15) -- (7.12,0) -- cycle    ;
\draw  [fill={rgb, 255:red, 0; green, 0; blue, 0 }  ,fill opacity=1 ] (374.8,130) .. controls (374.8,129.43) and (375.29,128.96) .. (375.9,128.96) .. controls (376.51,128.96) and (377,129.43) .. (377,130) .. controls (377,130.57) and (376.51,131.04) .. (375.9,131.04) .. controls (375.29,131.04) and (374.8,130.57) .. (374.8,130) -- cycle ;
\draw    (331,169.5) -- (351,170) ;

\draw (399.73,62.07) node [anchor=north west][inner sep=0.75pt]   [align=left] {\textcolor[rgb]{0,0.16,1}{{\fontfamily{ptm}\selectfont Controller}}};
\draw (97.33,129.27) node [anchor=north west][inner sep=0.75pt]    {$\mathbf{u}$};
\draw (134.8,71.8) node [anchor=north west][inner sep=0.75pt]  [font=\normalsize]  {$\mathbfcal{F}\mathbf{\left( u,x^{( 0)}\right)}$};
\draw (398.47,120.8) node [anchor=north west][inner sep=0.75pt]  [font=\normalsize]  {$\mathbfcal{M}_\theta\mathbf{( \cdot )}$};
\draw (137.83,23.3) node [anchor=north west][inner sep=0.75pt]    {$\mathbf{x}^{\mathbf{( 0)}}$};
\draw (226.7,63.7) node [anchor=north west][inner sep=0.75pt]    {$\mathbf{x}\boldsymbol{_{\ }}$};
\draw (474.23,108.1) node [anchor=north west][inner sep=0.75pt]    {$\mathbf{x}^{\mathbf{( 0)}}$};
\draw (60.4,37.93) node [anchor=north west][inner sep=0.75pt]   [align=left] {{\fontfamily{ptm}\selectfont \textcolor[rgb]{0.78,0,0}{System}}};
\draw (264.07,118.83) node [anchor=north west][inner sep=0.75pt]  [font=\normalsize]  {$\mathbf{PSF}_{\boldsymbol{\rho }}(\mathbf{x} ,\mathbf{u}_{\mathbf{L}})$};
\draw (367.23,108.1) node [anchor=north west][inner sep=0.75pt]    {$\mathbf{u}_{\mathbf{L}}$};
\draw (358.1,160.44) node [anchor=north west][inner sep=0.75pt]    {$\boldsymbol{\psi }(\mathbf{u}\mathrm{_{\mathbf{L}}})$};
\draw (315.1,153.73) node [anchor=north west][inner sep=0.75pt]    {$\boldsymbol{\rho }$};

\end{tikzpicture}

}
    \caption{The proposed framework. The PB operator $\mathbfcal M_\theta(\cdot)$ generates $\mathbf u_L$, which feeds both the scheduler $\boldsymbol{\psi}(\cdot)$ and the PSF. Using the resulting rate $\boldsymbol{\rho}$ and the state $\mathbf x$, the PSF filters $\mathbf u_L$ into the safe applied input $\mathbf u$.}
    \label{fig:new-framework}
\end{figure}

To address these limitations, we propose a control scheme decoupling safety, stability, and performance by integrating a parameterized PB controller with a scheduled PSF (Figure~\ref{fig:new-framework}).
Our contributions are: (i) a PSF whose Lyapunov-decrease rate is scheduled by the PB controller's $\ell_2$ input to reduce conservativeness while preserving recursive feasibility; (ii) theoretical guarantees of closed-loop $\ell_2$-stability and safety, proving that this architecture strictly expands the set of achievable trajectories compared to fixed-rate PSFs; and (iii) a data-driven actor-critic training procedure that bypasses the need for differentiating through the PSF optimization problem.


\subsection*{Notation}
The set of all sequences $\mathbf{x}=(x_0,x_1,\dots)$ with $x_t\in\mathbb{R}^n$ is denoted by $\ell^n$. The $t$-th element $x_t$ of a sequence $\mathbf{x}$ may also be denoted as $(\mathbf{x})_t$. For $p\in[1,\infty]$, define
$
\ell_p^n \;:=\; \{\mathbf{x}\in\ell^n \ |\ \|\mathbf{x}\|_{p}
  := (\sum_{t=0}^{\infty}|x_t|^p)^{1/p} < \infty\ \text{ for }p<\infty,\ 
  \|\mathbf{x}\|_{\infty}:=\sup_t |x_t| < \infty\},
$
for any vector norm $|\cdot|$.
We simplify $\|\cdot\|_2$ as $\| \cdot \|$. An operator $\mathbfcal{A}:\mathbf{x}\mapsto \mathbf{w}$ is said to be $\ell_2$-stable if it is causal and $\mathbfcal{A}(\mathbf{w})\in\ell_2^m$ for all $\mathbf{w}\in\ell_2^n$; equivalently, $\mathbfcal{A}\in\mathcal{L}_2$. We denote with $\mathbb{N}$ the set of positive integers and for some $a< b$, $\mathbb{N}_{[a,b]}=\{ n \in \mathbb{N} \ | \ a\leq n \leq b \}$. The prediction of a variable $x$ over a horizon of length $N\geq1$ at time $t$ are denoted by $x_{i|t}$ with $i\in\mathbb{N}_{[0:N-1]}$. The entire predicted trajectory is denoted by $x_{0:N|t}$, or, when clear from the context, by $x_{\cdot|t}$. Optimal quantities are marked with the superscript $^*$. The operator $\mathbb{E}_{x\sim\mathcal{D}}[\cdot]$ denotes expectation with respect to the random variable $x$ distributed according to $\mathcal{D}$.

\section{Problem Formulation}
Consider the \emph{noise-free} nonlinear discrete-time dynamical system:
\begin{equation} \label{eq: sys_eq}
    x_{t+1} = f(x_{t},u_{t}), \quad x_0\in\mathbb{R}^n,
\end{equation}
where $f:\mathbb{R}^n\times\mathbb{R}^m\rightarrow \mathbb{R}^n$, and $x_t \in \mathbb{R}^n$, $u_t \in \mathbb{R}^m$ are the state and control input, respectively.
The system~\eqref{eq: sys_eq} induces a unique causal transition map
\begin{equation}
\label{eq:sys_opt}
\mathbf{x} \;=\; \mathbfcal{F}(\mathbf{u},\mathbf{x^{(0)}}),
\end{equation}
where $\mathbfcal{F}: \ell^m\times\ell^n\rightarrow\ell^n$, $\mathbf{u}=(u_0,u_1,\dots)$, $\mathbf{x}=(x_0,x_1,\dots)$ and $\mathbf{x^{(0)}}:=(x_0,0,0,\dots)$ denotes the sequence that injects the initial condition.
For the system~\eqref{eq:sys_opt} equipped with a causal controller  $\mathbf{u}=\mathbfcal{K}(\mathbf{x})$, we denote the closed-loop mappings $\mathbf{x^{(0)}} \mapsto \mathbf{x}$ and $\mathbf{x^{(0)}} \mapsto \mathbf{u}$ as $\mathbf{\Phi}^\mathbf{x}(\mathbfcal{F}, \mathbfcal{K})$ and $\mathbf{\Phi}^\mathbf{u}(\mathbfcal{F}, \mathbfcal{K})$, respectively. Ideally, we wish the closed-loop system to be stable according to the following definition.
\begin{definition} \label{def:CLstab}
    For a given set $\mathcal{X}_s\subseteq\mathbb{R}^n$, the closed-loop system composed by~\eqref{eq:sys_opt} and the causal policy $\mathbf{u}=\mathbfcal{K}(\mathbf{x})$ is $\ell_2$-stable if $\mathbf{\Phi}^\mathbf{u}(\mathbfcal{F}, \mathbfcal{K})$, $\mathbf{\Phi}^\mathbf{x}(\mathbfcal{F}, \mathbfcal{K})$ belongs to $\mathcal{L}_2$ for all $x_0\in\mathcal{X}_s$.
\end{definition}
Moreover, the system~\eqref{eq: sys_eq} is subject to \emph{state and input constraints}, representing physical limitations and safe operating regions, which are expressed as $x_t \in \mathcal{X}$, $u_t \in \mathcal{U}$ for all $t\geq 0$, 
where $\mathcal{X}\subseteq\mathbb{R}^n$ and $\mathcal{U}\subseteq\mathbb{R}^m$ are nonempty, compact sets containing the origin. Beyond closed-loop stability and constraint satisfaction, we aim at \emph{optimizing performance}. We quantify performance by the infinite-horizon discounted loss
\begin{equation}
    L(\mathbf{x},\mathbf{u}) = \mathbb{E}_{x_0 \sim \mathcal{D}}\left[\sum_{t=0}^{\infty} \gamma^t l(x_t,u_t) \right],
\label{eq:obj_performance}
\end{equation}
where $l:\mathbb{R}^n \times \mathbb{R}^m \!\to\! \mathbb{R}$ is a piecewise differentiable stage cost and $0<\gamma<1$ is the discount factor. We assume $l$ is continuous on the compact set $\mathcal{X}\times\mathcal{U}$, which ensures the total loss $L(\mathbf{x},\mathbf{u})$ is bounded~\cite{bertsekas2012dynamic}.
The actual initial state is modeled as a random variable drawn from a distribution $\mathcal{D}$ with support $\mathcal{X}_s$. Consequently, the system evolves deterministically once $x_0$ is sampled, and the only source of uncertainty in closed-loop performance comes from the initial condition. 
\\
For guaranteeing stability, safety, and performance, our goal is to solve the following problem.
\begin{problem} \label{problem 1}
Find a nonlinear, time-varying
state-feedback controller $\mathbfcal{K}(\mathbf x)=(K_0(x_0),\dots,K_t(x_{0:t}),\dots)$ solving the following NOC problem.
   \begin{subequations} 
    \begin{align}
    \min_{\mathbfcal{K}(\cdot)} \quad & L(\mathbf{x},\mathbf{u}) \label{pb1: cost}\\
    \text{s.t.} \quad & x_{t+1} = f(x_{t},u_{t}),\label{pb1: dynamics constraints}\\
    & u_t = K_t(x_{0:t}), \label{pb1:control}\\
    & (\mathbf{\Phi}^\mathbf{x}(\mathbfcal{F}, \mathbfcal{K}), \mathbf{\Phi}^\mathbf{u}(\mathbfcal{F}, \mathbfcal{K})) \in \mathcal{L}_2, \label{pb1:stability}\\
    & x_{t} \in \mathcal{X}, u_{t} \in \mathcal{U}, \quad \forall t =0,1,\dots,\infty. \label{eq: pb1_con_states}
\end{align}
\end{subequations}
\end{problem}
Without the state and input constraints~\eqref{eq: pb1_con_states}, Problem~\ref{problem 1} is addressed in~\cite{furieri_mad_2025} via disturbance–feedback control.  This approach provides a complete parametrization of all and only the stability-preserving controllers for systems subject to an additive disturbance $\mathbf{w}\in\ell_p$, and it recasts the problem as learning an operator $\mathbfcal{M}\in\mathcal{L}_p$. In our noise-free setting, the role of $\mathbf{w}$ is played by the injected initial condition $\mathbf{x^{(0)}}$. 
However, that framework assumes $\mathcal{X}=\mathbb{R}^n$, $\mathcal{U}=\mathbb{R}^m$ and further, that a baseline (globally) stabilizing controller is already in place, making the closed-loop map from disturbances to states $\ell_p$-stable. As a result, it does not directly handle hard state and input constraints or plants that are only locally stabilized.


\section{Safe Performance Boosting}
\label{S:SafePB}
Our architecture (Fig. \ref{fig:new-framework}) decouples \emph{safety} (PSF-enforced state and input constraints), \emph{performance} (PB controller generating $\mathbf{u_L}$), and \emph{stability}. Stability is achieved synergistically: the PSF provides the Lyapunov certificate $J$, while the PB controller is parameterized to ensure $\mathbf{u_L} \in \ell_2$, eventually triggering a strictly stabilizing decrease in $J$. Below, we introduce the scheduled PSF and prove that this $\ell_2$ property guarantees closed-loop stability as per Definition \ref{def:CLstab}.

Throughout the paper, we work in the noise-free, $\ell_2$ setting. Once the inner loop, composed by the system and the PSF, is $\ell_2$-stable, an $\ell_2$ performance input $\mathbf{u_\mathrm{L}}$ (with a fixed initial condition $\mathbf{x^{(0)}}$) implies $(\mathbf{x,u})\in\ell_2$; hence trajectories are bounded (because $\ell_2 \subset \ell_\infty$). 
Extensions to other $p$-norms follow with minor modifications.

\subsection{PSF with scheduled Lyapunov-decrease}
We first introduce the PSF block in Figure~\ref{fig:new-framework}, which connects the operator $\boldsymbol{\mathcal{M}}(\cdot)$ and the plant $\mathbfcal{F}$. Given a prediction horizon $N\geq 1$, we define the following certificate function  $J:\mathbb{R}^{n\cdot(N+1)} \times \mathbb{R}^{m\cdot N}\rightarrow \mathbb{R}$
\begin{equation} \label{eq: lya_definition}
    J(x_{0:N},u_{0:N-1})= \sum_{i=0}^{N-1}s(x_i,u_i) + m(x_N),
\end{equation}
to serve as a Lyapunov function for closed-loop stability, where $s:\mathbb{R}^n \times \mathbb{R}^m \rightarrow \mathbb{R}_{\geq0}$ and $m:\mathbb{R}^n \rightarrow \mathbb{R}_{\geq0}$ are the stage and terminal functions, respectively.

The PSF aims to minimize the deviation between the filtered input and the performance input while ensuring constraint satisfaction and stability. Consider the following PSF with stability certificate~\cite{miliosStabilityMechanismsPredictive2024,didier2024predictive} at time $t$:
\begin{subequations} 
    \label{eq:PSF}
\begin{align}
    \min_{u_{\cdot|t}} \quad & \left\| u_{0|t} - u_{\mathrm{L},t} \right\|^2 \label{seq:psf-obj}\\
    \text{s.t.} \quad & x_{0|t} = x_t   \label{seq:psf-init}\\
    & x_{i+1|t} = f(x_{{i|t}},u_{{i|t}}), \quad \forall i \in \mathbb{N}_{[0,N-1]} \label{seq:psf-dya}\\
    & x_{i|t} \in \mathcal{X}, \quad \forall i \in \mathbb{N}_{[0,N-1]} \label{seq:psf-state-con}\\
    & u_{i|t} \in \mathcal{U}, \quad \forall i \in \mathbb{N}_{[0,N-1]}  \label{seq:psf-control-con}\\
    & x_{N|t} \in \mathcal{Z}_f \label{seq:psf-terminal-con} \\
    & \displaystyle J(x_{\cdot|t},u_{\cdot|t}) \leq  J(x^*_{\cdot|t-1},u^*_{\cdot|t-1}) \notag \\ & \qquad \qquad \qquad
    - (1-\rho)\cdot s(x^*_{0|t-1},u^*_{0|t-1}), \label{seq:stabconstr}
\end{align}
\end{subequations}
where $(x^*_{\cdot|t},u^*_{\cdot|t})$ denotes an optimizer of~\eqref{eq:PSF}, $\mathcal{Z}_f\subseteq\mathcal{U}$ is the terminal set and $u_{\mathrm{L},t} \in \mathbb{R}^m$ is the performance input from the PB controller. The first optimal predicted control input $u^*_{0|t}$ is applied to the system. Constraint~\eqref{seq:stabconstr} enforces a minimal decrease of $J$ at each time step. The parameter $\rho \in [0,1)$ sets the required decrease, guaranteeing a uniform stability margin $1-\rho>0$; smaller $\rho$ implies a stronger decrease \cite{didier2024predictive}. This yields a Lyapunov-like certificate. However, using a fixed decrease rate $\rho$ can shrink the admissible input set over time. We therefore modify~\eqref{seq:stabconstr} to reduce conservativeness during performance optimization. Specifically we relax the constraint $\rho\in[0,1)$ and allow a time-varying $\rho_t \in \mathbb{R}_{\ge 0}$ governed by a scalar function $\psi$.
\begin{definition}[Tightening schedule]\label{def:psi}
Fix $\bar\rho\in[0,1)$, $\varepsilon>0$ and a ceiling value $\rho_{\mathrm{max}}\geq 1$. The map $\psi:[0,\infty)\to[0,\infty)$ is a tightening schedule function if it is nondecreasing and continuous, and satisfy
\begin{align*}
    \text{(P1) } \psi(r)=\bar\rho\ \ &\forall r\in[0,\varepsilon], \qquad
\text{(P2) } \psi(r)\ge \bar\rho\ \ \forall r\ge 0, \\
& \text{(P3) } \psi(r)\le \rho_{\mathrm{max}}, \ \forall r\geq0.
\end{align*}
Given a performance input $u_{\mathrm{L},t}$, we set $\rho_t := \psi(\|u_{\mathrm{L},t}\|)$. This defines the causal operator $\boldsymbol{\psi}: \ell^m \to \ell$ that maps the signal $\mathbf{u_L}$ to the scheduling signal $\boldsymbol{\rho}$.
\end{definition}
The proposed tightening couples the performance input and the decreasing rate.
Indeed, if $\mathbf{u}_{\mathrm{L}}\in\ell_2$, then $\|u_{\mathrm{L},t}\|\to 0$. By (P1), there exists $T$ such that for all $t\ge T$ we have that $\|u_{\mathrm{L},t}\|\leq\epsilon$ and $\rho_t=\psi(\|u_{\mathrm{L},t}\|)=\bar\rho$. Thus, the schedule is automatically stabilizing for $t\geq T$ with uniform stability margin $1-\bar\rho>0$, while it can be more permissive in the transient $[0,T)$ (when $\|\mathbf{u}_{\mathrm{L}}\|$ is larger), reducing conservativeness.
A minimal schedule illustrating the ``if $\,\|u_{\mathrm{L},t}\|>\varepsilon$ then relax'' strategy is
\begin{equation} \label{eq:psifun}
    \psi(r)=
\begin{cases}
\bar\rho, & 
r\le \varepsilon,\\[2pt]
\bar \rho + (\rho_{\mathrm{max}}-\bar \rho) \min \big \{\frac{r-\varepsilon}{\varepsilon} ,1\big \}, & r> \varepsilon,
\end{cases}
\end{equation}
where $r:=\|u_{\mathrm{L},t}\|$. Note that prior stability-enhanced PSFs already allow time-varying Lyapunov-decrease rates (e.g., $\zeta_t$ in~\cite{miliosStabilityMechanismsPredictive2024}), but they are decoupled from the external controller providing $\mathbf{u}_{\mathrm{L}}$ and effectively bounded by $1$. Here, instead we couple the schedule $\rho_t$ to $u_{\mathrm L,t}$ and even permit $\rho_t>1$ transiently. This allows the PB controller to trade a transient Lyapunov-decrease for improved performance, while guaranteeing closed-loop stability.
With the proposed scheduler, at each time $t$ we solve~\eqref{eq:PSF} with~\eqref{seq:stabconstr} replaced by
\begin{subequations}    
    \begin{align}
    & \displaystyle J(x_{\cdot|t},u_{\cdot|t}) \leq J(x^*_{\cdot|t-1},u^*_{\cdot|t-1}) \notag\\
     & \qquad \qquad \quad- (1-\rho_t)\cdot s(x^*_{0|t-1},u^*_{0|t-1}). \label{eq: new-PSF-stability}
\end{align}
\end{subequations}
Let $(x^*_{\cdot|t},u^*_{\cdot|t})$ be an optimizer of \eqref{eq:PSF} and define the certificate optimal value by
$J_t^* := J(x^*_{\cdot|t},u^*_{\cdot|t})$. We apply the first input in receding-horizon fashion, $u_t := u^*_{0|t}$, inducing the PSF policy
$u_t = \kappa_{\mathrm{PSF}}(x_t, u_{\mathrm{L},t}, J^*_{t-1})$.
To initialize the PSF at $t=0$, we must set the initial value of the certificate cost $J_{-1}^*$. This is done by using any feasible warm start $(x^*_{\cdot|-1},u^*_{\cdot|-1})$ and setting
$J^*_{-1} := J(x^*_{\cdot|-1},u^*_{\cdot|-1})$; alternatively, one can set $J^*_{-1}=+\infty$ to deactivate
\eqref{eq: new-PSF-stability} at the first step, and repeat.

We introduce necessary assumptions to guarantee recursive feasibility and $\ell_2$-stability for the PSF~\eqref{eq:PSF}.

\begin{assumption} \label{ap:cost function}
The stage cost $s:\mathbb{R}^n\times\mathbb{R}^m\to\mathbb{R}_{\ge 0}$ is continuous with $s(0,0)=0$ and is positive definite on $\mathcal X\times \mathcal U$, i.e., $s(x,u)>0$ for all $(x,u)\in (\mathcal X\times \mathcal U)\setminus\{(0,0)\}$. Moreover, there exists constants $q_x,q_u>0$ such that
\[
s(x, u) \;\ge\; q_x\|x\|^2 \;+\; q_u\|u\|^2, \qquad \forall (x,u) \in \mathcal{X} \times \mathcal{U}.
\]
\end{assumption}
\begin{assumption} \label{ap: terminal set}
There exist a terminal set $\mathcal{Z}_f \subseteq \mathcal{X}$, a terminal cost $m: \mathcal{Z}_f \to \mathbb{R}_{\ge 0}$, and an auxiliary control law $\kappa: \mathcal{Z}_f \to \mathcal{U}$ such that:

\begin{enumerate}[label=(\roman*)]
    \item $\kappa(x) \in \mathcal{U}$ and $f(x,\kappa(x)) \in \mathcal{Z}_f$;
    \item $ m(f(x,\kappa(x))) - m(x) \le -\,s(x, \kappa(x)) $.
\end{enumerate}
\end{assumption}
These assumptions are standard in MPC settings. Assumption~\ref{ap:cost function} provides a quadratic lower bound that, allows one to prove that states and inputs are in $\ell_2$, and is met, e.g., by a quadratic stage cost in \eqref{eq: lya_definition}. The next result precises this statement. 
Assumption~\ref{ap: terminal set} requires a pre-stabilizing controller in the terminal set $\mathcal{Z}_f$. 
A classical way to satisfy it is to use linearized terminal ingredients (see e.g.,~\cite{chen1998quasi}); a simpler alternative, altough more restrictive in terms of region of attraction~\cite{rawlings2020model}, is to impose a terminal equality in place of a terminal set, i.e., $\mathcal Z_f=\{0\}$ and $m\equiv 0$. 

Before stating the main theorem, we highlight its key non-trivial aspect. Standard PSF stability proofs~\cite{didier2024predictive,miliosStabilityMechanismsPredictive2024} rely on a uniform Lyapunov decrease, requiring the decrease rate $\rho$ to be strictly less than $1$ at all times. Our analysis is fundamentally different: the decrease rate $\rho_t = \psi(\|\mathbf{u_L}\|)$ is signal-dependent and explicitly allowed to be $\ge 1$ transiently. This invalidates any proof based on a per-step decrease. Instead, our theorem proves $\ell_2$-stability by showing that the signal property $\mathbf{u_L} \in \ell_2$ is sufficient to eventually enforce the uniform decrease rate $\bar\rho$, ensuring closed-loop $\ell_2$-stability.

\begin{theorem}\label{prop:ell2_stability}
Suppose Assumptions~\ref{ap:cost function}--\ref{ap: terminal set} hold and let the PSF~\eqref{eq:PSF} use the scheduled stability constraint \eqref{eq: new-PSF-stability} where $\rho_t = \psi(\|u_{\mathrm{L},t}\|)$ is generated by a tightening schedule $\psi$ satisfying Definition~\ref{def:psi}. If this problem is feasible at time $t=0$, then it is feasible at all $t\ge 1$. Moreover, if 
$\mathbf{u_\mathrm{L}}\in\ell_2$, then the induced closed loop is $\ell_2$-stable, i.e., $\mathbf{x},\mathbf{u}\in\ell_2$, and $x_t\in\mathcal{X}$, $u_t\in\mathcal{U}$, $\forall t\geq 0$.
\end{theorem}
The proof can be found in Appendix~\ref{app:thm1}.

\subsection{Behavioral characterization and strict dominance in $\ell_2$}
We characterize the closed-loop trajectories generated by the PSF when coupled with the PB controller. By Theorem~\ref{prop:ell2_stability}, once~\eqref{eq:PSF} with~\eqref{eq: new-PSF-stability} is feasible at $t=0$, the PSF remains feasible and the closed-loop is $\ell_2$-stable for any performance input $\mathbf{u_\mathrm{L}}\in\ell_2$. Hence, the key question is how the set of $\ell_2$ closed-loop trajectories realizable by our scheduled architecture compares to that of a fixed-decrease PSF \eqref{eq:PSF}. We prove our architecture's set (i) contains and (ii) strictly enlarges the fixed-decrease set.
\\
We define a \emph{tightening profile} as any sequence $\boldsymbol\rho=\{\rho_t\}_{t\ge0}$ with $\rho_t\in[0,\rho_{\mathrm{max}})$. Let $\mathcal{P}$ be a set of tightening profiles. The set of all possible $\ell_2$ closed-loop trajectories induced by $\mathcal{P}$ is defined as
\begin{align}\label{eq:behavior_family}
\mathcal B(\mathcal P)
:= & \big\{\,(\mathbf x, \mathbf u)\in\ell_2^n\times\ell_2^m\ \big|\
 \exists\,x_0\in\mathcal X,\ \exists\,\mathbf{u_\mathrm{L}}\in\ell_2,\ \notag  \\
&  \quad \exists\,\boldsymbol\rho\in\mathcal P:\notag\text{~\eqref{eq:PSF} with~\eqref{eq: new-PSF-stability} is feasible at $t=0$, } \notag \\ & \quad \text{and yields }(\mathbf x,\mathbf u)\big\}. \notag
\end{align}
We will compare the sets of achievable trajectories induced by the following sets
\begin{align}
\mathcal P_{\mathrm{fix}}(\bar\rho)&:=\{\boldsymbol\rho:\ \rho_t\equiv\bar\rho\},\\
\mathcal P_{\mathrm{sch}}(\psi)&:=\big\{\boldsymbol\rho:\ \exists\,\mathbf{u_\mathrm{L}}\in\ell_2 \text{ s.t. } \rho_t=\psi(\|u_{\mathrm{L},t}\|)\big\}.
\end{align}
The overall set of trajectories $\mathcal{B}$ is determined by the admissible input set the PSF allows at each time step, which in turn depends on the tightening profile $\mathbf{\rho}$. We define this admissible input set and the induced one-step reachable set as follows:
\begin{definition}\label{def:U_t}
At time $t$, given $x_t$, the certificate $J^*_{t-1}$ from the previous time, and $\rho\in[0,\rho_{\mathrm{max}})$, define
\begin{align*}
\mathcal U_t(\rho,J) :=& \Big\{\,  v\in\mathcal U\ \Big|\ 
 \exists\ (x_{1:N|t},u_{1:N-1|t})\ \text{s.t.}\\
& x_{0|t}=x_t,\ u_{0|t}=v, \  \eqref{seq:psf-dya}-\eqref{seq:psf-terminal-con}, \\
& J(x_{\cdot|t},u_{\cdot|t})\le J-(1-\rho)\,s(x^*_{0|t-1},u^*_{0|t-1})\ \Big\},
\end{align*}
where $J:=J^*_{t-1}$ and $(x^*_{\cdot|t-1},u^*_{\cdot|t-1})$ is the previous optimizer.
The one-step reachable set $\ \mathcal X^+_t(\rho,J):=\{\,f(x_t,v):\ v\in\mathcal U_t(\rho,J)\,\}$.
\end{definition}

\begin{lemma}
\label{lem:monotone_Ut}
Fix $x_t$ and $(x^*_{\cdot|t-1},u^*_{\cdot|t-1})$. If $\rho_1\leq \rho_2$ and $J_1\leq J_2$, then
\[
\mathcal U_t(\rho_1,J_1) \subseteq \mathcal U_t(\rho_2,J_2)\ ,
\qquad
\mathcal X^+_t(\rho_1,J_1) \subseteq \mathcal X^+_t(\rho_2,J_2).
\]
Moreover, if $s(x^*_{0|t-1},u^*_{0|t-1})>0$ and the constraint~\eqref{seq:stabconstr} is active at $(\rho_1,J_1)$, then \(
\mathcal U_t(\rho_1,J_1) \subsetneq \mathcal U_t(\rho_2,J_2)
\)
whenever either $\rho_2>\rho_1$ or $J_2>J_1$.
\end{lemma}
The proof can be found in Appendix~\ref{app:lem1}.

We now lift the per-step result from Lemma~\ref{lem:monotone_Ut} to compare the full sets of achievable trajectories. We will show that the set from the scheduled architecture strictly contains all trajectories achievable by the fixed-$\rho$ architecture.

\begin{theorem}\label{thm:inclusion_strict}
Fix an initial value of the certificate function $J^*_{-1}$. Let $\bar\rho\in[0,1)$ and let $\psi$ be a tightening schedule. Then
\begin{equation}\label{eq:incl_main}
\mathcal B\!\left(\mathcal P_{\mathrm{fix}}(\bar\rho)\right)
\ \subseteq\
\mathcal B\!\left(\mathcal P_{\mathrm{sch}}(\psi)\right).
\end{equation}
Moreover, suppose there exists a closed-loop trajectory under the scheduled architecture from some $x_0\in\mathcal X$ and a time $\tilde {t}$ such that
\begin{equation}\label{eq:strict_cond}
\rho_t^{\mathrm{sch}}>\bar\rho
\quad\text{and}\quad
\mathcal U_{\tilde{t}}(\rho_{\tilde{t}}^{\mathrm{sch}},J^*_{\tilde{t}-1})\ \supsetneq\
\mathcal U_{\tilde{t}}(\bar\rho,J^*_{\tilde{t}-1}).
\end{equation}
Then there exists a performance input $\tilde{\mathbf u}_\mathrm{L}\in\ell_2$ for which the scheduled PSF produces a trajectory
\begin{equation} \label{eq:strictSeparation}
    (\mathbf x,\mathbf u)\ \in\
\mathcal B\!\left(\mathcal P_{\mathrm{sch}}(\psi)\right)\setminus
\mathcal B\!\left(\mathcal P_{\mathrm{fix}}(\bar\rho)\right).
\end{equation}
\end{theorem}
\smallskip
The proof can be found in Appendix~\ref{app:thm2}.\\
\smallskip
Theorem~\ref{thm:inclusion_strict} shows that a fixed-rate PSF constrains the applied control action $u_t$ to the set $\mathcal U_t(\bar\rho,J^*_{t-1})$. The proposed scheduled architecture only enforces this stability constraint ($\rho_t = \bar\rho < 1$) after the transient (i.e., once $\|u_{\mathrm{L},t}\| \le \varepsilon$) and relaxes the bound otherwise. Consequently, whenever $\psi(\|u_{\mathrm{L},t}\|)>\bar\rho$ and the Lyapunov inequality is active, the admissible input set $\mathcal U_t$ strictly enlarges. The proof of Theorem~\ref{thm:inclusion_strict} shows that applying any input from this enlarged set-difference at even a single time step results in a closed-loop trajectory that is unattainable by the fixed-rate architecture.

\subsection{Parameterization of the PB controller}\label{sec:learner-param}
\noindent
By Theorem~\ref{prop:ell2_stability}, closed-loop $\ell_2$-stability and constraint satisfaction are guaranteed provided the performance input $\mathbf{u_\mathrm{L}}\in\ell_2$ for any $x_0 \in \mathcal{X}_s$, where $\mathcal{X}_s=\{x\in\mathcal{X}: \text{Problem~\eqref{eq:PSF} with~\eqref{eq: new-PSF-stability} is feasible at } t=0 \}$. This section describes a parametrization for the PB controller $\mathbfcal{M}$, that ensures this $\ell_2$ condition on $\mathbf{u_L}$ holds by construction.

We adopt the Magnitude-and-Direction (MAD) policy of~\cite{furieri_mad_2025} to parameterize the operator $\mathbfcal{M}_\theta(\mathbf{x},\mathbf{x^{(0)}})$ as
\begin{align}
\label{eq:MAD}
\mathbf{u}_{\mathrm{L}}
=\mathbfcal{M}_\theta(\mathbf{x},\mathbf{x^{(0)}})
=
|\mathbfcal{A}(\mathbf{x^{(0)}})| \odot \mathbfcal{D}(\mathbf{x}),
\end{align}
where $\odot$ denotes the elementwise product, $\mathbfcal{A}\in\mathcal{L}_2$, and $\mathbfcal{D}(\mathbf{x})\in\ell_\infty^m$ for all $\mathbf{x}\in\ell^n$. For instance, $\mathbfcal{D}$ can be induced by the bounded static map $d(x_t)=\tanh(\mathrm{NN}_{\theta_1}(x_t))\in[-1,1]^m$. Since $\|\mathbfcal{D}(\mathbf{x})\|_\infty\le 1$ and $\mathbfcal{A}\in\mathcal{L}_2$, the resulting input $\mathbf{u}_{\mathrm{L}}\in\ell_2$ by construction for all $\theta$~\cite{furieri_mad_2025}.
To parameterize the nonlinear operator $\mathbfcal{A}\in\mathcal{L}_2$, we use a Linear Recurrent Unit (LRU)~\cite{orvieto2023resurrecting}, one of several learning architectures that guarantee $\mathbfcal{A}(\cdot,\theta_2)\in\mathcal{L}_2$ for all $\theta_2\in\mathbb{R}^{n_{\theta_2}}$~\cite{revayRecurrentEquilibriumNetworks2023,orvieto2023resurrecting,massai2025free,zakwan2024neural}. Consequently, the full MAD policy guarantees $\mathbf{u_L}\in\ell_2$ for all $\theta=[\theta_1^\top,\theta_2^\top]^\top\in\mathbb{R}^{n{\theta_1}+n_{\theta_2}}$.

\subsection{Training procedure}
Direct end-to-end differentiation through the PSF is challenging because the PSF is defined by a parametric optimization problem whose solution map may become non-differentiable when active sets change~\cite{andersson_sensitivity_2018}. 
We therefore treat the PSF--plant interconnection as a black-box augmented system and train $\mathbfcal{M}_\theta$ with an off-policy actor--critic method, e.g., DDPG~\cite{lillicrap2015continuous}, following~\cite{furieri_mad_2025}.
Specifically, we optimize the discounted infinite-horizon performance loss with the actor $\mu_\theta$ defining the learning input $u_{\mathrm{L},t}=\mu_\theta(\zeta_t)$.
Using deterministic policy gradients (DPG)~\cite{silver_deterministic_nodate}, the actor gradient is
\begin{equation}
\label{eq:policy-gradient}
\nabla_{\theta} L(\theta)
\;=\; 
\mathbb{E}_{\zeta_t\sim \rho^{\mu}}\!\left[
\nabla_{\theta} \mu_\theta(\zeta_t)\;
\nabla_{u_\mathrm{L}} Q_\mu(x_t,u_\mathrm{L})\big|_{u_\mathrm{L}=\mu_\theta(\zeta_t)}
\right],
\end{equation}
where $\zeta_t$ is the input to the actor\footnote{The definitiom of $\zeta_t$ is problem-specific. For the recurrent operator in Section~\ref{sec:learner-param}, it includes the physical state $x_t$, the element from the initial condition sequence $(\mathbf{x^{(0)}})_t$, the internal state of the recurrent network, and the previous value of the certificate $J^*_{t-1}$.} and the expectation $\mathbb{E}_{\zeta_t\sim \rho^{\mu}}$ is taken over the distribution of actor inputs $\zeta_t$ visited under the policy $\mu$.
The critic $Q_\mu(x, u_\mathrm{L})$ estimates the value of applying the learning input $u_\mathrm{L}$ when the system state is $x$. It is trained to minimize the error in satisfying the Bellman equation. In our noise-free setting, the PSF-plant dynamics are deterministic: a given $(x, u_\mathrm{L})$ pair produces a single, unique next state $x^+ = f(x, \kappa_{\mathrm{PSF}}(x,u_\mathrm{L},J^*))$. This simplifies the Bellman equation to:
\begin{multline}
\label{eq:bellman-deterministic}
Q_\mu(x,u_\mathrm{L}) = l(x, \kappa_{\mathrm{PSF}}(x,u_\mathrm{L},J^*)) \\ + \gamma Q_\mu(x^+, \mu_\theta(\zeta^+))
\end{multline}
where $\zeta^+$ is the next actor input corresponding to $x^+$. This deterministic form is simpler than the standard Bellman equation for stochastic systems, as it does not require an expectation over the next state $x^+$~\cite{lillicrap2015continuous,silver_deterministic_nodate}.
Crucially, the gradient calculation in~\eqref{eq:policy-gradient} does not require the derivatives of the PSF solution map $\kappa_{\mathrm{PSF}}$. Instead, the gradient of the $Q$-function with respect to $u_\mathrm{L}$ (which is learned by the critic) serves as the necessary gradient information, bypassing the problematic end-to-end differentiation. 
During training, exploration is added to $u_{\mathrm{L},t}$, and the PSF filters unsafe proposals, preserving safety and stability throughout data collection.

\section{Numerical Example}
\label{sec:numsection}

\begin{figure}[!t]
     \centering
    \includegraphics[width=0.9\columnwidth]{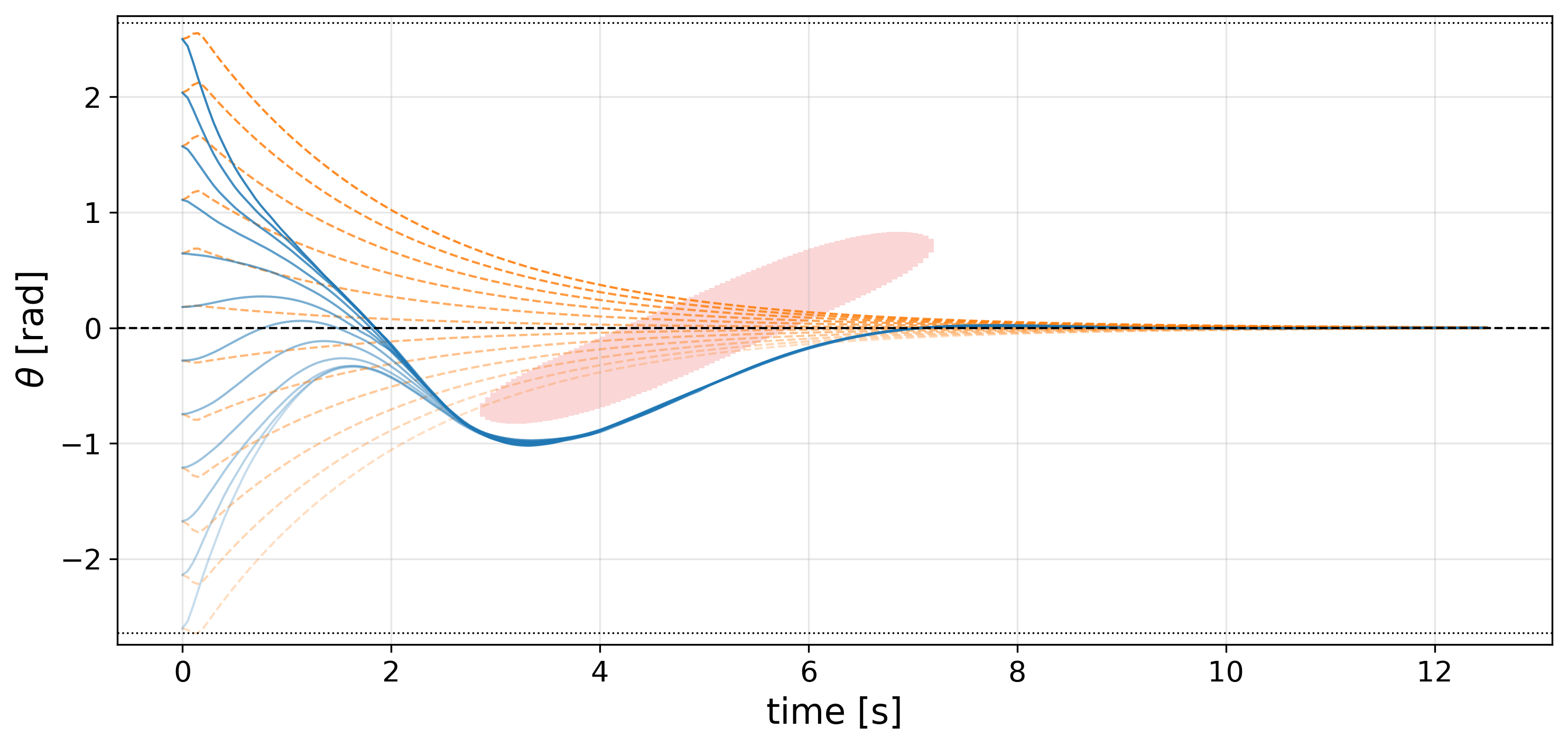}%
    \caption{Angle trajectories with obstacle avoidance. Lines with varying opacity correspond to multiple initial conditions. Blue lines: the proposed approach. Orange lines: the PSF with $\rho_t\equiv\bar\rho$ and $\mathbf{u_\mathrm{L}}\equiv \mathbf{0}$. The translucent red region shows the moving obstacle in the $(t,\theta)$ plane; the dotted black line marks the lower and upper bounds, while the dashed black line marks the unstable equilibrium ($\theta=0$).
 }
    \label{fig:theta}
\end{figure}
In this section, we demonstrate the proposed scheme on a pendulum stabilization task chosen to highlight a setting where alternative methods can fail. The goal is to stabilize the pendulum at its unstable upright equilibrium, which is not directly compatible with IMC-based approaches requiring a stable or pre-stabilized plant~\cite{furieriLearningBoostPerformance2024,furieri_mad_2025}. The task also includes a moving obstacle, forcing a detour in the angular trajectory. As we show, this detour requires a transient increase of the Lyapunov certificate $J^*$, which is unattainable under a fixed-rate PSF but enabled by our scheduling mechanism.

We consider the pendulum dynamics:
\begin{align*}
    \dot \theta = \omega, \quad
    \dot \omega = -\frac{g}{l}\sin{\theta} - \frac{b}{ml^2}\dot{\theta} + \frac{1}{ml^2}u,
\end{align*}
where $\theta$ and $\omega$ denote the angle and angular velocity, respectively; $m=0.2\,\mathrm{Kg}$ is the mass; $l=0.5\,\mathrm{m}$ is the pendulum length; and $g$ and $b=0.02\,\mathrm{N\, m\, s/rad}$ represent the gravitational acceleration and damping coefficient.  We obtain a discrete time model using a fourth-order Runge--Kutta (RK4) step with sampling time $T_s=0.05$ $\mathrm{s}$.
The safe sets and bounds are $\theta \in [-2.6,2.6] \ \mathrm{rad}$, $u \in [-3,3] \mathrm{Nm}$, and the task is to stabilize the pendulum upright to the unstable equilibrium ($\theta=0$) while avoiding a moving obstacle with center in $p^{\mathrm{obs}}_t$ that sweeps at a fixed speed $v^{\mathrm{obs}}=0.2$ $\mathrm{m/s}$ from right to left on the $x$ axis. In Figure~\ref{fig:theta}, the translucent red region shows the space-time area occupied by the moving obstacle.
As in Section~\ref{S:SafePB}, the inner loop is the PSF with horizon $N=20$. For this example, we use a terminal equality constraint ($x_{N|t}=0$) and define the Lyapunov certificate $J$ from \eqref{eq: lya_definition} using the stage cost $s(x_k, u_k) = \|x_k\|_{Q}^{2}+\|u_k\|_{R}^{2}$ ($Q\succeq0$, $R\succ0$) and a terminal cost of zero ($m \equiv 0$). The stability constraint uses the tightening schedule $\rho_t = \psi(\| u_{\mathrm{L},t} \|)$ with parameters $\bar{\rho}=0.5$, $\varepsilon=0.05$, and $\rho_{\mathrm{max}}=10$ (see Definition~\ref{def:psi}), and the function $\psi$ is a smooth version of~\eqref{eq:psifun}. 
The MAD operator $\mathbf{u_\mathrm{L}}=\mathbfcal{M}_\theta (\mathbf{x},\mathbf{x^{(0)}})$ parametrizes $\mathbfcal{A}\in\mathcal{L}_2$ using an LRU. 
Since the task involves a moving obstacle, the performance objective~\eqref{eq:obj_performance} becomes the time-varying, discounted loss: $L_{\mathrm{tv}}(\theta)= \mathbb{E}_{x_0\sim\mathcal D}\Big[\sum_{t\ge 0}\gamma^t  l_t(x_t,u_t)\Big],$ where the time-varying stage cost $l_t$ is defined as the weighted sum:
\begin{multline*}
    l_t(x_t,u_t, u_{\mathrm{L},t}) = \beta_1 l_{\mathrm{traj}}(x_t,u_t) + \beta_2 l_{\mathrm{psf}}(u_t,u_{\mathrm{L},t}) \\ + \beta_3 l_{\mathrm{obs},t}(x_t),
\end{multline*}
where $\beta_i > 0$ are positive weights. Here, $l_{\mathrm{traj}}$ is a quadratic tracking and control-effort penalty, $l_{\mathrm{obs},t}$ penalizes collisions with the time-varying obstacle, and $l_{\mathrm{psf}}$ adds a small penalty on PSF intervention. 
The moving obstacle induces a time-varying stage cost, which we handle through a standard augmented-state reformulation $\tilde x_t := [ x_t^\top , c_t^\top ]^\top$, where $c_t := [p^{\mathrm{obs}^\top}_t, v^{\mathrm{obs}^\top}]$ contains the obstacle's state. 
The PSF constraints, terminal ingredients, and Lyapunov schedule $\boldsymbol{\psi}(\cdot)$ still act only on the physical state $x_t$, so the proofs of recursive feasibility and eventual $\ell_2$-stability remain identical. We only extend the MAD policy's input to $\mathbf{u}_{\mathrm L}=\mathbfcal M_\theta(\tilde{\mathbf{x}},\mathbf{x}^{(0)})$, which still guarantees $\mathbf u_{\mathrm L}\in\ell_2$ by construction (see Sec.~\ref{sec:learner-param}).
We implemented the approach in PyTorch\footnote{Code to reproduce the experiments and figures, as well as an animated visualization of the task, is available at  \url{https://github.com/DecodEPFL/PSF_PB.git}.}.
We compare the scheduled architecture against a fixed-rate PSF with $\rho_t=\bar{\rho}$ and $\mathbf{u_{L}}\equiv \mathbf{0}$. This baseline enforces a monotonic decrease in $J$, confining trajectories to shrinking level sets and thus forbidding transient detours. By Theorem~\ref{thm:inclusion_strict}, our architecture recovers all fixed-rate trajectories when $\mathbf{u_{L}} \equiv \mathbf{0}$ and can strictly enlarge this set. For a fair comparison, both methods use the same certificate $J$, horizon $N$, and stability margin $\bar\rho$. Unsafe RL baselines are omitted, as they provide no safety guarantees.
\begin{figure}[t]
     \centering
    \includegraphics[width=0.9\columnwidth]{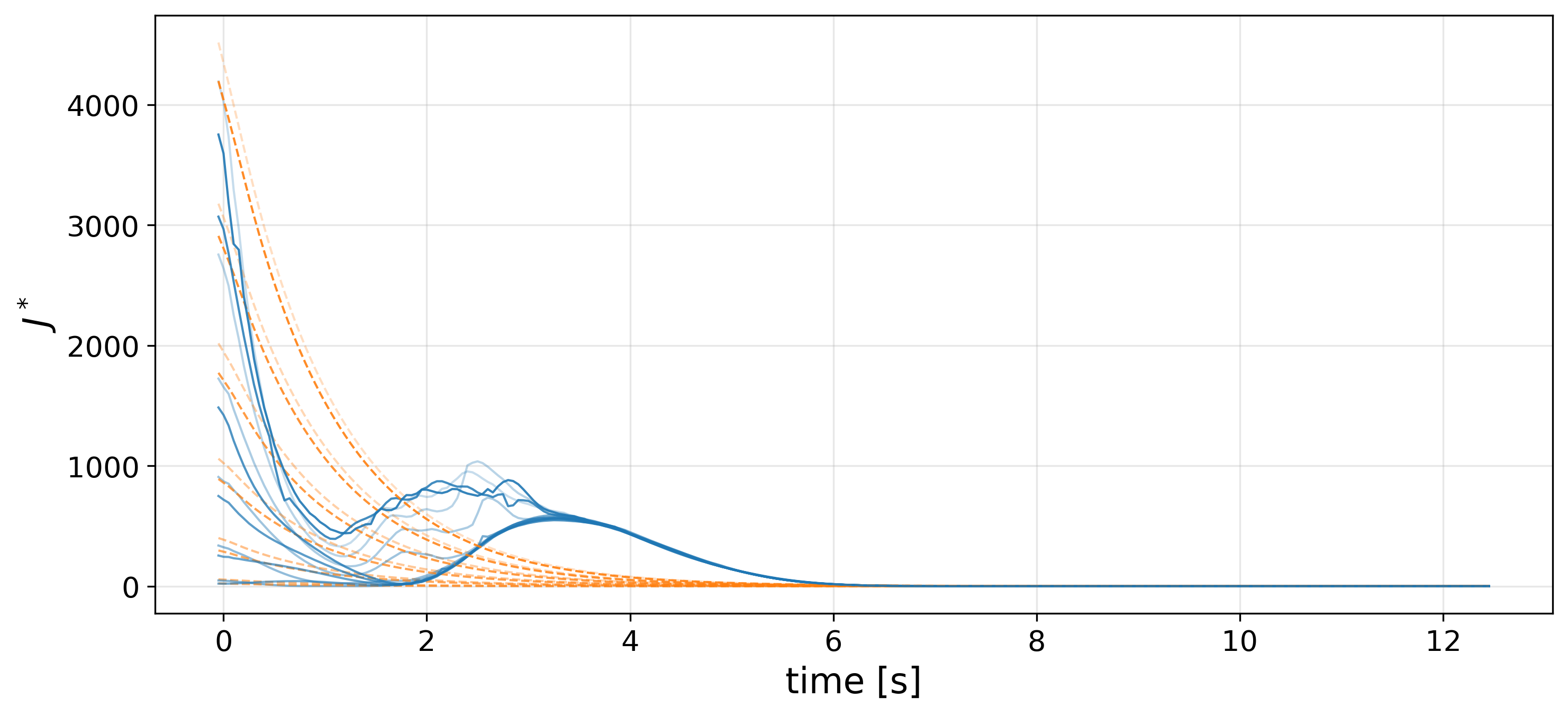}%
    \caption{Time evolution of the PSF certificate $J_t^*$ for different initial conditions. Blue lines: the proposed approach. Orange lines: the PSF with $\rho_t\equiv\bar\rho$ and $\mathbf{u_\mathrm{L}}\equiv \mathbf{0}$.}
    \label{fig:J}
\end{figure}
Figures~\ref{fig:theta} and \ref{fig:J} show the angle $\theta_t$ and Lyapunov certificate $J^*_t$ trajectories. Both methods respect the safety bounds (Fig.~\ref{fig:theta}, dotted lines). However, the fixed-$\bar \rho$ baseline (orange) fails the task, as its path would collide with the obstacle (red region). This failure is fundamental: Fig.~\ref{fig:J} (orange) shows $J^*_t$ is forced to decrease monotonically, trapping the system in shrinking level sets and making the detour provably impossible. Our scheduled architecture (blue) succeeds by breaking this limitation. As shown in Fig.~\ref{fig:J} (blue), $J^*_t$ is allowed to increase transiently ($\rho_t > \bar\rho$), providing the freedom to execute the safe detour (Fig.~\ref{fig:theta}, blue). After bypassing the obstacle, $\mathbf{u_L} \to 0$, $\rho_t$ tightens to $\bar\rho$, and $J^*_t$ converges to zero, ensuring eventual stability. This behavior illustrates the per-step set enlargement $\mathcal{U}_t(\rho,J)$ and the strict separation shown in Theorem~\ref{thm:inclusion_strict}.

\section{Conclusions}
This paper introduced a control architecture integrating a PB controller with a scheduled PSF to decouple safety and performance. We proved that scheduling the Lyapunov-decrease rate strictly expands the set of achievable safe trajectories, enabling transient detours as demonstrated in a pendulum obstacle-avoidance task. The proposed actor-critic training method successfully avoids differentiating through the PSF's optimization map. Future work includes extending the stability analysis to $\ell_p$-gains, incorporating state estimation, and generalizing the approach to multi-agent coordination.


\bibliographystyle{IEEEtran}
\bibliography{references}    

\newpage

\appendix
\subsection{Proof of Theorem~\ref{prop:ell2_stability}}
\label{app:thm1}

We start by proving the recursive feasibility of the PSF. Assume the problem~\eqref{eq:PSF} with~\eqref{eq: new-PSF-stability} is feasible at time $t-1$. We define the candidate solution at time $t$ by using standard shifting arguments~\cite{rawlings2020model}: for $i=0,\dots,N-2$, set
$\hat{x}_{i|t}=x^*_{i+1|t-1}$ and $\hat{u}_{i|t}=u^*_{i+1|t-1}$, $\hat{u}_{N-1|t}=\kappa(x^*_{N|t-1})$ and
$\hat{x}_{N|t}=f(x^*_{N|t-1},\kappa(x^*_{N|t-1}))$. By
Assumption~\ref{ap: terminal set}(i), state/input constraints and the terminal set condition hold. For the certificate function, consider the following candidate
\begin{multline*}
J(\hat{x}_{\cdot|t},\hat{u}_{\cdot|t})
= \sum_{i=1}^{N-1} s(x^*_{i|t-1},u^*_{i|t-1})
     \\
     \qquad +s(x^*_{N|t-1},\kappa(x^*_{N|t-1})) + m\!\big(f(x^*_{N|t-1},\kappa(x^*_{N|t-1}))\big),
\end{multline*}
using Assumption~\ref{ap: terminal set}(ii) we have
\begin{align*}
J(\hat{x}_{\cdot|t},\hat{u}_{\cdot|t})&\le \sum_{i=1}^{N-1} s(x^*_{i|t-1},u^*_{i|t-1}) + m(x^*_{N|t-1})  \\
&= J(x^*_{\cdot|t-1},u^*_{\cdot|t-1}) - s(x^*_{0|t-1},u^*_{0|t-1}) \\
&\le J(x^*_{\cdot|t-1},u^*_{\cdot|t-1}) - (1-\rho_t)\, s(x^*_{0|t-1},u^*_{0|t-1}).
\end{align*}
Since \(s(\cdot)\!\ge\!0\), the inequality holds for $0\leq \rho_t<1$ (and is trivially looser for \(\rho_t\ge1\)); hence the shifted candidate is feasible at time \(t\). 
Since the PSF~\eqref{eq:PSF} with~\eqref{eq: new-PSF-stability} is recursively feasible and because the optimizer satisfies \eqref{eq: new-PSF-stability}, we have $J(x^*_{\cdot|t},u^*_{\cdot|t})\leq J(\hat x_{\cdot|t},\hat u_{\cdot|t})$, obtaining:
\begin{equation}
\label{eq:perStepCost}
J_t
\;\le\;
J_{t-1}
\;-\;
(1-\rho_t)\, s_{t-1} .
\end{equation}
with $J_t=J(x^*_{\cdot|t},u^*_{\cdot|t})$ and $s_{t-1}=s(x^*_{0|t-1},u^*_{0|t-1})$.
Because $\mathbf{u}_\mathrm{L}\in\ell_2$ we have $\|u_{\mathrm{L},t}\|\to 0$. By Definition~\ref{def:psi}(P1), there exists $T$ such that for all $t\ge T$ we have $\rho_t=\psi(\|u_{\mathrm{L},t}\|)=\bar\rho$. Define the uniform margin $\underline{\eta} \;:=\; 1-\bar\rho \;>\; 0$.
Summing \eqref{eq:perStepCost} from $t=T+1$ to $k$ yields
\begin{align*}
    \sum_{t=T+1}^k (J_{t-1}-J_t) &\geq \sum_{t=T+1}^k (1-\rho_t) s_{t-1}, \\
    J_T-J_k \geq \sum_{t=T+1}^k (1-\rho_t) &s_{t-1} \geq \underline{\eta}\sum_{t=T+1}^k s_{t-1}
\end{align*}
Since $s(\cdot), \ m(\cdot)\geq0$ one has $J_k\geq0$ and, hence
\begin{equation*}
    \underline{\eta}\sum_{t=T+1}^{k} s_{t-1}
\;\le\; J_T-J_k\;\le\;
J_T.
\end{equation*}
By using Assumption~\ref{ap:cost function} and letting $k\rightarrow\infty$,
\[
\underline{\eta}\sum_{i=T}^{\infty} \big(q_x\|x^*_{0|i}\|^2 + q_u\|u^*_{0|i}\|^2\big)
\;\le\;
J(x_{\cdot|T},u_{\cdot|T})
\;<\;\infty,
\]
so that $\sum_{t=0}^\infty \|x_t\|^2<\infty$ and $\sum_{t=0}^\infty \|u_t\|^2<\infty$, i.e., $\boldsymbol{x},\boldsymbol{u}\in\ell_2$. 
Constraint satisfaction at all times follows from recursive feasibility.
\hfill $\blacksquare$

\subsection{Proof of Lemma~\ref{lem:monotone_Ut}} \label{app:lem1}
The pair $(\rho,J)$ only enters in the definition of $\mathcal{U}_t$ through
\(
J(x_{\cdot|t},u_{\cdot|t})\le J-(1-\rho)\,s(x^*_{0|t-1},u^*_{0|t-1}).
\)
Increasing $\rho$ or $J$ relaxes the Lyapunov constraint while all other constraints are unchanged. Thus, any plan feasible for $(\rho_1,J_1)$ remains feasible for $(\rho_2,J_2)$, proving the inclusions. 
Moreover, if the Lyapunov inequality is active and $s(\cdot)>0$, this relaxation strictly enlarges the set: $
\mathcal U_t(\rho_1,J_1) \subsetneq \mathcal U_t(\rho_2,J_2)
$. 
\hfill $\blacksquare$

\subsection{Proof of Theorem~\ref{thm:inclusion_strict}} \label{app:thm2}
We first prove the inclusion~\eqref{eq:incl_main}. 
Any trajectory $(\mathbf x,\mathbf u) \in \mathcal B(\mathcal P_{\mathrm{fix}}(\bar\rho))$ can be reproduced by the scheduled architecture by setting the performance input $u_{\mathrm{L},t}:=u_t$. Since $\rho_t^{\mathrm{sch}}=\psi(\|u_t\|)\ge\bar\rho$, by Lemma~\ref{lem:monotone_Ut}, $u_t$ remains a feasible action. As $u_{0|t}=u_t$ achieves zero cost for the PSF objective \eqref{seq:psf-obj}, it is optimal.
Proceding by induction on $t$ yields \eqref{eq:incl_main}. 

We now prove the strict separation~\eqref{eq:strictSeparation}.
Pick any $v\in \mathcal U_{\tilde {t}}(\rho_{\tilde {t}}^{\mathrm{sch}},J^*_{\tilde{t}-1})\setminus \mathcal U_{\tilde {t}}(\bar\rho,J^*_{\tilde {t}-1})$ given \eqref{eq:strict_cond}.
Construct a performance input $\tilde{\mathbf u}_L\in\ell_2$ verifying $\tilde u_\mathrm{L,\tau}=u_\tau$ for $\tau<\tilde {t}$, $\tilde u_{\mathrm{L},\tilde {t}}=v$ and $\tilde u_\mathrm{L,\tau}=0$ for $\tau>\tilde {t}$.
Up to time $\tilde {t}-1$ both architectures can apply the same inputs (by~\eqref{eq:incl_main}), so they share the same $J^*_{\tilde {t}-1}$.
The scheduled PSF can apply $u_{\tilde{t}}=v$ (its optimal action), while the fixed-rate PSF cannot (as $v$ is infeasible for it). The resulting trajectory is therefore in the set difference, proving \eqref{eq:strictSeparation}.
\hfill $\blacksquare$

\end{document}